\documentclass[journal,onecolumn]{IEEEtran}

\usepackage{graphicx}
\usepackage{subfigure}
\usepackage{wrapfig}
\usepackage{enumerate}
\usepackage{enumitem}
\usepackage{color}
\usepackage{wrapfig}
\usepackage{algorithm}
\usepackage{algpseudocode}
\usepackage{mathtools}
\usepackage{eucal}
\usepackage{mathptmx}

\usepackage{booktabs}
\usepackage{cite}
\usepackage{amssymb}
\usepackage{algorithmicx}
\usepackage{algorithm,caption}
\usepackage{algpseudocode}
\usepackage{color,soul}
\usepackage{multicol}

\ifCLASSINFOpdf
\else
\fi
\begin{document}

\title{Complexity-based Financial Stress Evaluation}

\author{Hongjian Xiao, Yao Lei Xu and Danilo P. Mandic}

\maketitle

\begin{abstract}
Financial markets typically exhibit dynamically complex properties as they undergo continuous interactions with economic and environmental factors. The Efficient Market Hypothesis indicates a rich difference in the structural complexity of security prices between normal (stable markets) and abnormal (financial crises) situations. Considering the analogy between market undulation of price time series and physical stress of bio-signals, we investigate whether stress indices in bio-systems can be adopted and modified so as to measure 'standard stress' in financial markets. This is achieved by employing structural complexity analysis, based on variants of univariate and multivariate sample entropy,
to estimate the stress level of both financial markets on the whole and the performance of the individual financial indices. Further, we propose a novel graphical framework to establish the sensitivity of individual assets and stock markets to financial crises. This is achieved through Catastrophe Theory and entropy-based stress evaluations indicating the unique performance of each index/individual stock in response to different crises. Four major indices and four individual equities with gold prices are considered over the past 32 years from 1991-2021. Our findings based on nonlinear analyses and the proposed framework support the Efficient Market Hypothesis and reveal the relations among economic indices and within each price time series. 

\end{abstract}


\IEEEpeerreviewmaketitle

\section{Introduction}

Over the last several decades, structural Complexity characteristics of stock markets have been investigated owing to their potential efficient indicators associated with financial health and economic stabilization  \cite{Ref182,Ref189}. The cornerstone of modern financial theory, the Efficient Market Hypothesis (EMH), states that the underlying value of an asset incorporates all the available information to make the stock always trade at a fair value \cite{Ref257}. This implies that in 'normal' situations, markets are barely influenced by any specific event and evenly respond to the continuous stimuli of economic change. In the context of complexity science, the security prices in 'normal' situations exhibit high randomness and uncertainty, together with low predictability from their historical values. On the other hand, when an economic crisis occurs, corresponding to an 'abnormal' situation, markets are easily driven by irrational fluctuations (e.g. panic buying/selling), showing a lower degree of randomness and higher determinism \cite{Ref98,Ref99}.

Analogous to the well-known automatic 'fight-or-flight' response in physiological stress studies from human cognitive science \cite{Ref256}, financial stress can also be considered as a deviation from the normal functioning of financial markets \cite{Ref252}. Considering the similarity between the Complexity Loss Theory (CLT) in human body functions \cite{Ref254} and the implications of EMH in financial markets, the concept of sympatho-vagal balance in bio-systems can also be used to describe the 'acceleration-stabilization' type of behaviour in financial systems \cite{Ref102, Ref141}. In human-centred sciences, the sympatho-vagal balance refers to the joint influence of the sympathetic nervous system (SNS) and parasympathetic nervous system (PNS) \cite{Ref102}, which respectively accelerates and decelerates body functions. Regarding the analysis of financial systems, the acceleration-stabilization' behaviour is driven and sustained by supply and demand, which results in market expansions and recessions \cite{Ref98}. 

By virtue of the analogy between financial systems and bio-systems, the non-parametric analyses, that have been applied in human-centred science to reveal complex features in the nonlinear domain, can also be employed to estimate and determine the financial stress of markets \cite{Ref57, Ref58, Ref129, Ref208}. Nonlinear methods utilized in the following analyses include entropy-based univariate/multivariate algorithms. Examples include the Recurrence Quantification Analysis (RQA) \cite{Ref183} and Assessment of Latent Index of Stress using instantaneous amplitude (iA-ALIS) methods \cite{Ref141}. Traditional entropy methods aiming at signal irregularity quantification have been applied in financial investigations due to their model-independent and prejudice-free analysis, including Approximate Entropy \cite{Ref57, Ref129}, Sample Entropy \cite{Ref69}, Permutation Entropy \cite{Ref341, Ref342} and other recently developed algorithms of a kind \cite{Ref363, Ref364, Ref444}. In addition to univariate entropy analysis, enhanced multivariate methodologies have been initially involved such as Multivariate Multiscale Sample Entropy \cite{Ref147}. Recurrence Quantification Analysis (RQA) is another popular nonlinear measure method in quantifying determinism via predictability, which has been widely considered in physiological studies \cite{Ref152, Ref416}. When it comes to financial systems, stock market volatility is a well-established concept referring to asset price fluctuations and reflects the degrees of uncertainty of future prices \cite{Ref57, Ref99, Ref129}. Closely relevant to the volatility measure, the high degree of determinism (DET) given by RQA implies small volatility and high predictability \cite{Ref181, Ref190}. Assessment of latent index of stress (ALIS) was introduced to detect financial crises primarily in stock markets by examining the power in Low-Frequency and High-Frequency bands of dynamical prices \cite{Ref98, Ref99}. Expanding on the ALIS, an assessment based on the instantaneous amplitude (iA-ALIS) was proposed in place of the power of the signal as a more reliable indicator of financial stress \cite{Ref141}.

As stated in EMH, the asset price contains all available information, and in order to obtain the dynamics of price change, most financial studies have been implemented based on the return time series, the difference between the two consecutive prices, $x(t+1)$ and $x(t)$. In this work, to maintain maximal information of price time series in the signal, we first apply a Moving Average (MA) filter to detrend data. Given the potential of nonlinear methods to reveal financial stress, we employ Complexity Science (CS), that is structural complexity analysis based on historical data to predict the occurrence of an 'abnormal' situation, that is, a financial crash. Such a sudden change in the behaviour of the financial system results from the smooth changes which arise jointly from both economic and non-economic factors. This phenomenon has been conceptualized by Ren{\'e} Thom, who termed it Catastrophe Theory \cite{Ref481}. While applicable to any dynamical system, the use of Catastrophe Theory in economics has been limited to a small number of studies \cite{Ref448, Ref449, Ref451, Ref452}. We therefore set out to demonstrate that the approaches of financial stress measurement offer a very insightful and practical way for quantitative analysis. To this end, we considered 32 years of historical data, from January 1991 to December 2021, with one data point per weekday. The main investigation is conducted over 4 stock indices (Dow Jones Industrial Average, NASDAQ Composite, Russell 2000 and Standard \& Poor's 500) and 4 individual equities from different industries (Apple Inc., Microsoft, McDonald's and American International Group). The Catastrophe plots exhibit the unique performance of each index/individual stock in response to different crises in the last several decades.

The remainder of the paper is organized as follows. Details of nonlinear algorithms and methodologies are presented in Section~\ref{sec.alg}. To give an overview of the asset price time series and analytical flow, a summary of data and methods is given in Section~\ref{sec.data}. Section~\ref{sec.result} illustrates the results of measures and analyzes the nonlinear properties of financial markets. Section~\ref{sec.cata} proposes the framework of Catastrophe Theory in financial investigation and provides initial analyses based on Catastrophe Theory. Finally, Conclusions are given in the last~section.

\section{Algorithm and Methods}
\label{sec.alg}
Algorithms used in this study are first introduced and described here.

\subsection{Modified Univariate Multiscale Sample Entropy \& Modified Multivariate Multiscale Sample Entropy}

\subsubsection{Moving Average filter}
To obtain the scaled and detrended signal, $y^{(\tau)}(j)$, a Moving Average (MA) filter is first utilized to remove the local trend, $s^{(\tau)}(j)$, from the original time series, $\{x(i)\}_{i=1}^N$, as
\begin{equation}
\label{eq.MA_1}
                s^{(\tau)}(j) = \frac{1}{\tau}\sum^{j+\tau/2-1}_{i = j-\tau/2 - 1}{x(i)},\quad 1\leq j\leq N-\tau+1.
\end{equation}

\begin{equation}
\label{eq.MA_2}
                 y^{(\tau)}(j)  = x(i) - s^{(\tau)}(j)
\end{equation}

\noindent where $\tau$ is a pre-defined scale factor. Observe that, different from the traditional Coarse Graining Process \cite{Ref23}, the scaling given by the MA filter is capable of maintaining the original signal length and is a better fit to the intrinsic properties of the applied data.

\subsubsection{Multivariate Sample Entropy}

Sample Entropy is a standard approach to evaluate the irregularity and randomness degree of time series based on their temporal dynamics, which has been widely applied in real-world complex systems \cite{Ref22, Ref26}. Sample Entropy is built based on the probability of similarity between embedding (delay) vectors, where their higher similarity demonstrates the higher predictability at multiple scales. To this end, we applied the MA filter to implement the scaling process for each time series, which was termed the Modified Multiscale Entropy (Mod-MSE) \cite{Ref97}. Due to the intrinsic properties of financial data, the pre-defined scale factor is set as one week, $\tau = 5$. The Mod-MSE was applied on every index and stock individually to give the stress evaluation of each index and equity, thus indicating their response towards the external environment. However, any individual stock or single index is not sufficient to represent the performance of the whole financial market. To this end, the enhanced Multivariate Entropy is employed to estimate the overall stress of the financial market. The multivariate entropy method accounts for the cross-channel dependencies in multivariate data by constructing Composite Delay Vectors (CDV), $\textbf{X}_M(i)$, derived from the original $p$-channel signal, $\{x_{k,i}\}_{i=1}^N,\,1\leq k\leq p$, in the form

  \begin{align*}
                \textbf{X}_M(i) =[& x_{1,i},x_{1,i+l_1},\dots,x_{1,i+(m_1-1)l_1)},\\
                                 & x_{2,i},x_{2,i+l_2},\dots,x_{2,i+(m_2-1)l_2)},\\
                                 &\qquad \qquad \qquad \vdots \\
                                 & x_{p,i},x_{p,i+l_p},\dots,x_{p,i+(m_p-1)l_p)},]
    \label{eq.CDV}\tag{3}
    \end{align*}
where $m_k$ and $l_k$ denote respectively the embedding dimension and time delay set to  $k^{th}$ channel.

Based on a combination of the MA filter (as a scaling process) with Multivariate Sample Entropy, the Mod-MMSE was employed across multiple channels as a standard stress estimation method over the whole market, with the details of Mod-MSE and Mod-MMSE given in Algorithm \ref{alg.Mod-MMSE}.

\begin{algorithm}[htb]

    \caption{Modified Univariate Multiscale Sample Entropy (Mod-MSE) \& Modified Multivariate Multiscale Sample Entropy (Mod-MMSE)}
    \label{alg.Mod-MMSE}
    \begin{algorithmic}[0]
    \Statex Given a multivariate data set with $P$ channels $\{x_{k,i}\}_{i=1}^N,\,1\leq k\leq p$, of length $N$, or a univariate data set with $P=1$.
    
    \begin{enumerate}
        \item Standardize the original data sets by subtracting the mean and dividing by the standard deviation for each channel.
        
        \item Scale the normalized datasets, $\{y_{k,i}^{(\tau)}\}_{j=1}^{N-\tau+1}$, for each channel following on (\ref{eq.MA_1}) and (\ref{eq.MA_2}).
        
        \item Form the Composite Delay matrix, $\textbf{Y}_M(i)$, according to the embedding dimension, $M$, and the time delay, $L$, as in Equation (\ref{eq.CDV}).

        \item Compute the distance between all the pairwise Composite Delay Vectors, $Y_M(i)$ and $Y_M(j)$, based on the Chebyshev distance, as $d_M(i,j) = max\{ Y_M(i+k)-Y_M(j+k)||\, i \neq j\}$. The number of matching patterns, $B_M(i)$, is defined as the similar pairs of delay vectors that satisfy the criterion $d_M(i,j)\leq r$.
        
        \item Compute the estimated local probability of $B_{M}(i)$ by $C_M(i) = \frac{B_{M}(i)}{N-n-1}$, where $n = max(M)*max(L)$, and the estimated global probability is $\Phi_{M}=\frac{\sum_{i=1}^{N-n}{{C_{M}(i)}}}{N-n}$.
      
        \item Repeat Steps 1 - 5 with an increased embedding dimension, $M^*=M+1$, and obtain the updated global probability, denoted as $\Phi_{M^*}=\frac{\sum_{i=1}^{N-n}{{C_{M^*}(i)}}}{N-n},\, n = max(M^*)*max(L)$.
        
           \item Modified Univariate/Multivariate Multiscale Sample Entropy is defined as
           \begin{equation}
               Mod\mbox{-}MMSE(m,l,r,N) = -\ln{[\frac{\Phi_{M^*}}{\Phi_M}]}.
               \nonumber
           \end{equation}
           
    \end{enumerate}

    \end{algorithmic}
\end{algorithm}

\subsection{Recurrence Quantification Analysis}
The Recurrence Plot (RP) is a traditional methodology to identify hidden correlations in multi-dimensional spaces, without the limitation of data stationarity and size restriction \cite{Ref79, Ref135}. By using Takens' embedding theorem \cite{Ref259}, the univariate time series, $\{x(i)\}_{i=1}^N$, is reconstructed into a phase space according to the optimal embedding dimension, $m$, and time delay, $l$, as
\begin{equation}
\label{eq.EM}
 x_m(i) = [x(i),x(i+l),\dots,x(i+(m-1)l)]
 \end{equation}
 
The optimal combination of embedding dimension and time delay can be selected via different methods, such as False Nearest Neighbours for embedding dimension \cite{Ref179} and Minimum Mutual information for delay factor \cite{Ref139}. In this work, we choose the joint selection of optimal embedding dimension and time delay by the differential entropy-based method introduced in \cite{Ref136}.
 
\begin{algorithm}[htb]

    \caption{Recurrence Quantification Analysis (RQA)}
    \label{alg.RQA}
    \begin{algorithmic}[0]
    \Statex Given a univariate data set $\{x(i)\}_{i=1}^N$ of length $N$.
    
    \begin{enumerate}
         \item Construct the delay vectors (DVs), $x_m$, according to Takens' embedding theorem, as in Equation (\ref{eq.EM}).
        
        \item Generate the Recurrence Plot (RP) matrix, composed of pairwise Euclidean distances between DVs, as
         \begin{equation}
              RP(i,j) = \Theta(\varepsilon-||x_m(i)-x_m(j)||,\, i,j = 1,\dots, N-n-1,\,i\neq j,
                \nonumber
            \end{equation}
        where $||\cdot||$ designates the Euclidean distance, $\Theta(\cdot)$ refers to the Heaviside function, and $\varepsilon$ denotes the threshold when defining the similarity between DVs, which is set as 60\% of the mean Euclidean distance of the DVs, and $n = (m-1)*l$.
        
        \item The degree of Determinism (DET) can be calculated as the percentage of recurrence points that form diagonal lines in the RP matrix, that is
        \begin{equation}
              DET = \frac{\sum^{N-n-1}_{j = j_{min}}j\cdot P(j)}{\sum^{N-n-1}_{j = 1}j\cdot P(j)}
                \nonumber
            \end{equation}
            where $P(j)$ is the number of diagonal lines of the length $j$, and $j_{min}$ is the minimal number of points to be considered as a diagonal line, which is set as $j_{min}=2$.
            
    \end{enumerate}

    \end{algorithmic}
\end{algorithm}

The outcome of RP is a matrix summarizing the distance among the Delay Vectors (DVs). Given a threshold, $\varepsilon$, the RP matrix can be plotted as a grey image, where every element in the matrix is converted into a pixel colour based on the relation between $\varepsilon$ and the distances between DVs. Several probabilistic measures can be implemented according to the RP matrix, such as Degree of Determinism (DET) and Laminarity (LAM). Both could be used as measures of the inverse of the volatility, where DET is the percentage of recurrent points forming diagonal line structures and LAM is that of forming vertical lines \cite{Ref181}. In financial markets, volatility is an important property that shows the implicit risk and is generally referred to as the degree of uncertainty about the future price \cite{Ref129}, hence, it also reflects the degree of predictability. Here, we employed the DET index in RQA, where the length of a diagonal line in RP reflects the number of consecutive recurrent states. Considering the inverse relation of DET and volatility, we expect that high determinism refers to the high predictability of the future price, representing the 'abnormal' situation in the financial market \cite{Ref332}. To this end, the change of DET is positively related to the stress level of the estimated index. Algorithm \ref{alg.RQA} outlines the process of DET calculation.

\subsection{Assessment of Latent Index of Stress with Instantaneous Amplitude (iA-ALIS)}

Assessment of Latent Index of Stress (ALIS) was especially proposed to quantify the 'stress level of a financial organism' in \cite{Ref99}. The original ALIS utilised the low-frequency band power and high-frequency band power of detrended price after the MA filter. By taking into account the intrinsic properties of financial data, the Low-Frequency band is considered to occupy the frequency band below 0.0042Hz ( =$\frac{1}{240}$) with a time window of one year, while the High-Frequency band is set to between 0.0167Hz ( =$\frac{1}{60}$) and 0.2Hz ( =$\frac{1}{5}$), corresponding to 2 months and 5 days, respectively. Recently, by employing instantaneous amplitude via the Hilbert transform, ALIS has been enhanced with correct signal power estimation, as discussed in \cite{Ref141}. We therefore employed the Assessment of Latent Index of Stress with Instantaneous Amplitude (iA-ALIS) based on the detrended financial time series. The higher the iA-ALIS index the more stressful the stock, with the threshold between 'stressed' and 'normal' derived based on the median value. The details of iA-ALIS are summarised in Algorithm \ref{alg.ALIS}.

\begin{algorithm}[htb]
    \caption{Assessment of Latent Index of Stress with Instantaneous Amplitude (iA-ALIS)}
    \label{alg.ALIS}
    \begin{algorithmic}[0]
    \Statex Given a univariate data set $\{x(i)\}_{i=1}^N$ of length $N$.
    
    \begin{enumerate}
         \item Remove the trend of the data by a Moving Average filter with a window of 1 year, and obtain the detrended data, $\{z(i)\}_{i=1}^N$.
         
         \item Bandpass-filter the detrended data, $z(i)$, into the Low-Frequency (LF) Band and High-Frequency (HF) bands.
         
         \item Apply the Hilbert transform to LF and HF, and obtain the instantaneous amplitude based on the analytic signals given by the Hilbert transform at every time point, denoted as $iA_{LF}$ and $iA_{HF}$. 
         
         \item Take 4 years as the window length and 1 month as an increment. In every time window, exclude the 20\% largest and smallest values, to remove the outliers and then calculate the mean $iA$ for every window, denoted as $LF(d)$ and $HF(d)$, where $d$ refers to a month.
         
         \item Normalise the $LF(d)$ and $HF(d)$ series by subtracting the mean and dividing by standard deviation in order to alleviate the problem of scaling.
         
         \item Remove the offset of $LF(d)$ and $HF(d)$.
         
         \item The ALIS is given by $ALIS(d) = LF(d) +HF(d)$
         
         \item The median value in the $ALIS(d)$ is a threshold for the stress in the market.
    \end{enumerate}

    \end{algorithmic}
\end{algorithm}

\section{Data Overview and Methods Summary}
\label{sec.data}

We applied several methodologies to four groups of indices/stocks over the last 32 years, spanning the period between 1991.01.01 and 2021.12.31. These are:

\vspace{0.3cm}

\begin{minipage}[htb]{0.58\linewidth}
  \begin{itemize}
    \item Stock market index
    \begin{itemize}
    \item Dow Jones Industrial Average (DJIA/DOW): 30 large companies;
	\item NASDAQ Composite (NAS): Mid- and large-caps;
	\item Russell 2000 (RUS): Smaller companies;
	\item Standard \& Poor's 500 (SNP): 500 large companies.
    \end{itemize}
    
    \item Equity
    \begin{itemize}
    \item Apple Inc.: Large technology company;
	\item Microsoft: Large technology company;
	\item McDonald’s: Fast food company;
	\item American International Group (AIG): Insurance~company.
    \end{itemize}
 \end{itemize}     
\end{minipage}
\hfill
\begin{minipage}[htb]{0.38\linewidth}
  \begin{itemize}
     \item Price of Metal
    \begin{itemize}
    \item Gold (Au);
	\item Silver (Ag);
	\item Copper (Cu);
	\item Platinum (Pt).
    \end{itemize}
    
     \item Currency
    \begin{itemize}
    \item EUR-GBP; 
	\item GBP-JPY;
	\item GBP-USD;
	\item USD-JPY.
    \end{itemize}
\end{itemize}
\end{minipage}

\vspace{0.3cm}

\begin{figure}[H]
\begin{minipage}[b]{.48\linewidth}
  \centerline{\includegraphics[width=\linewidth]{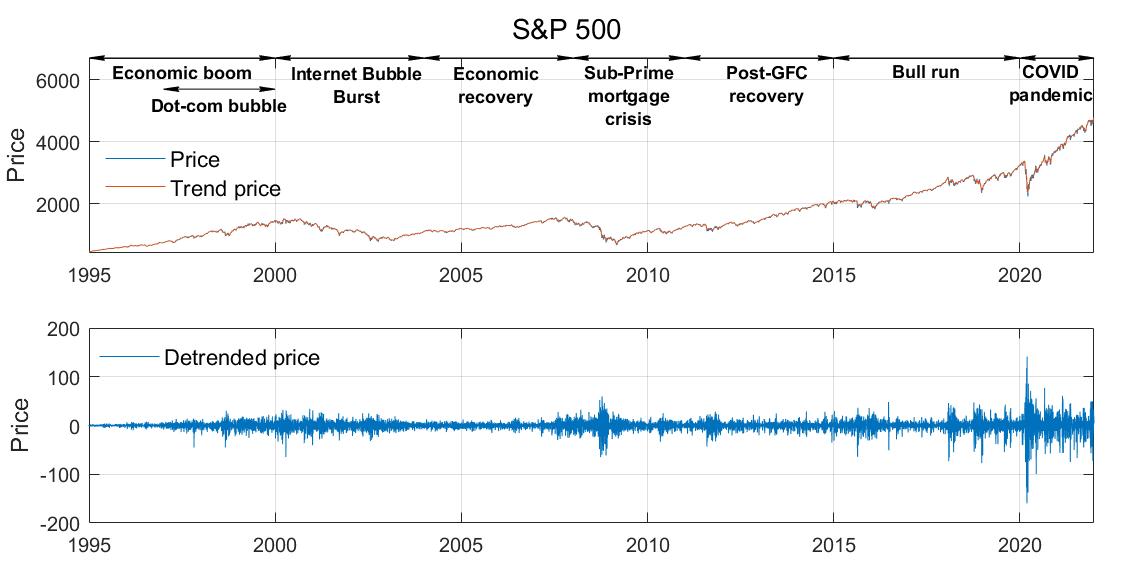}}
\end{minipage}
\hfill
\begin{minipage}[b]{0.48\linewidth}
  \centerline{\includegraphics[width=\linewidth]{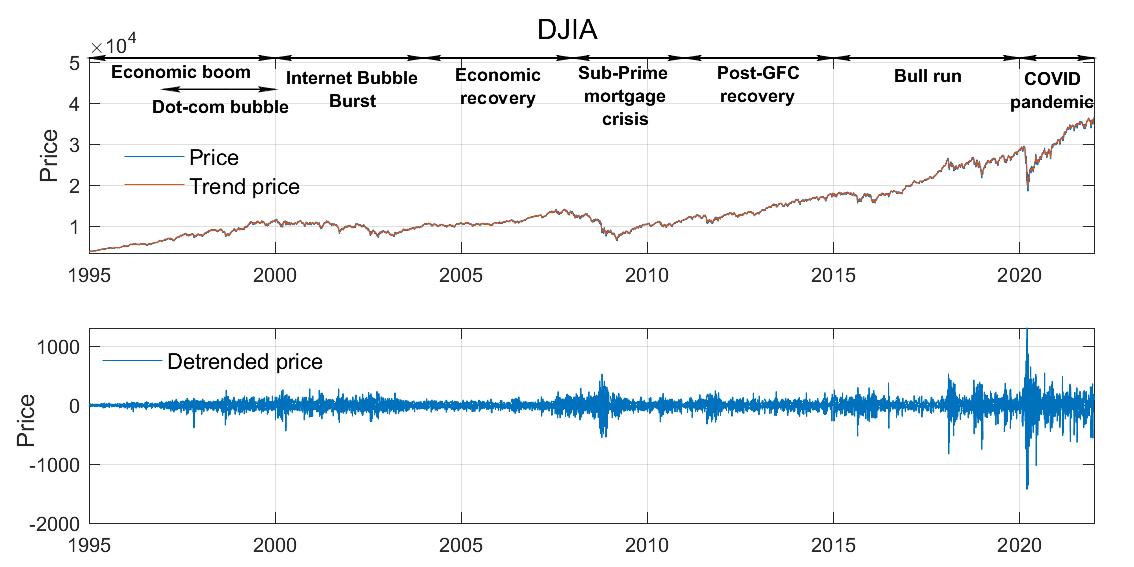}}
\end{minipage}

\begin{minipage}[b]{.48\linewidth}
\centerline{\includegraphics[width=\linewidth]{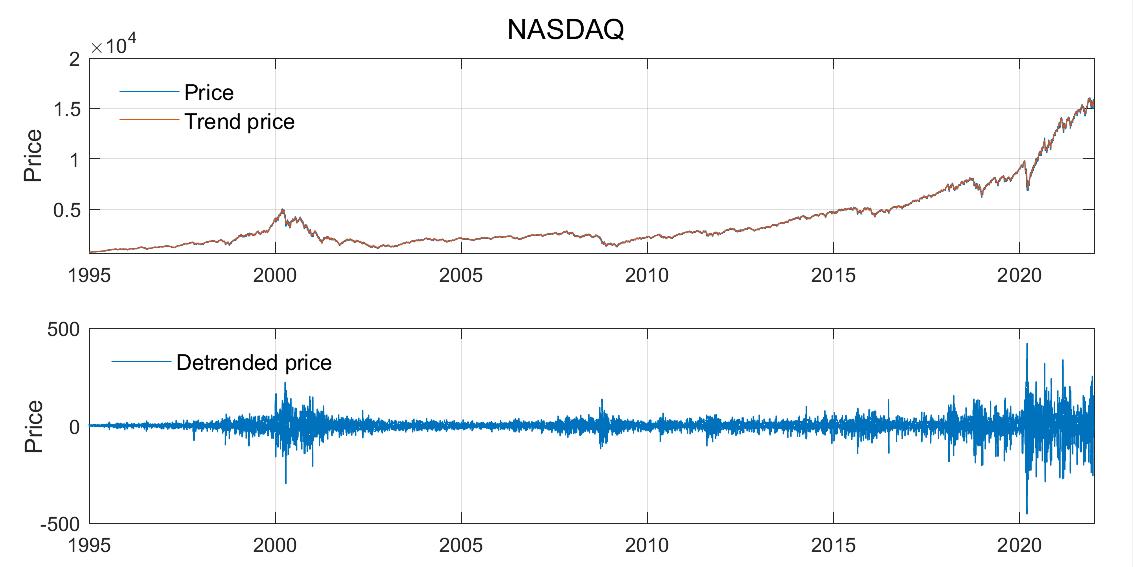}}
\end{minipage}
\hfill
\begin{minipage}[b]{0.48\linewidth}
  \centerline{\includegraphics[width=\linewidth]{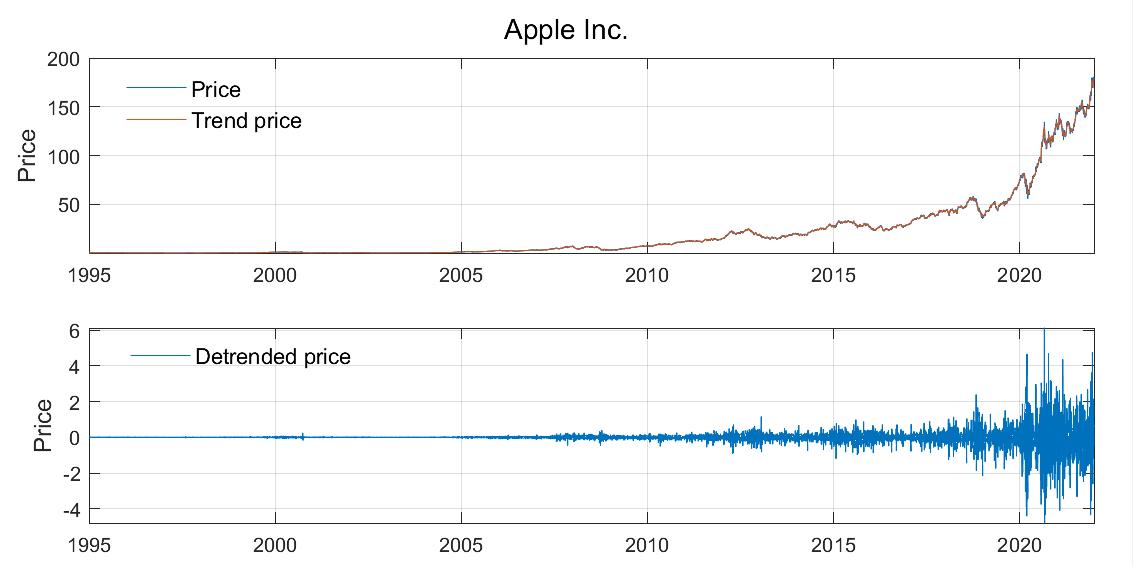}}
\end{minipage}

\begin{minipage}[b]{.48\linewidth}
  \centerline{\includegraphics[width=\linewidth]{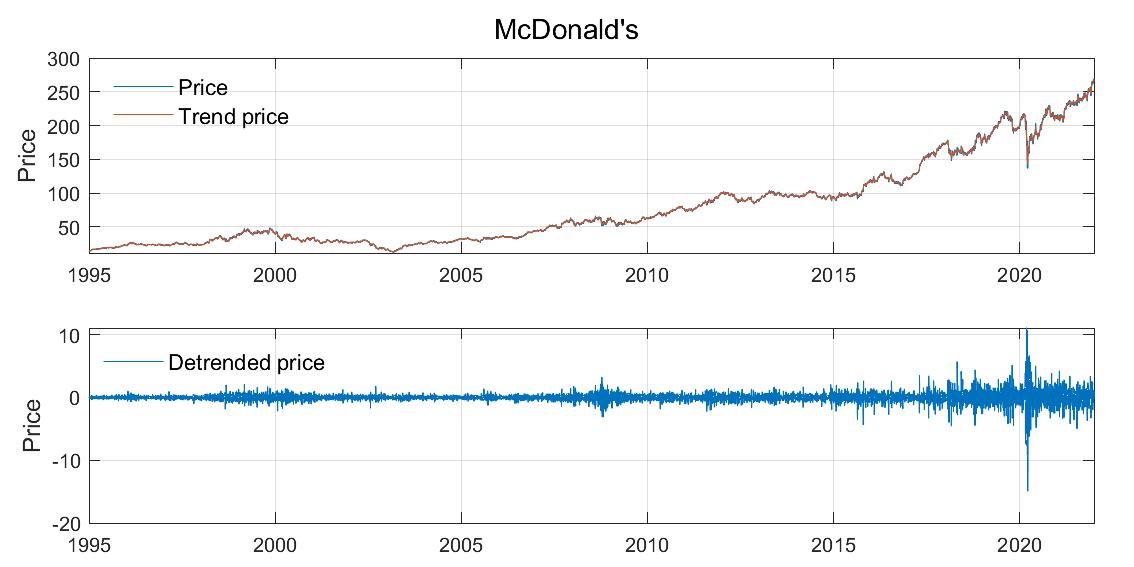}}
\end{minipage}
\hfill
\begin{minipage}[b]{0.48\linewidth}
\centerline{\includegraphics[width=\linewidth]{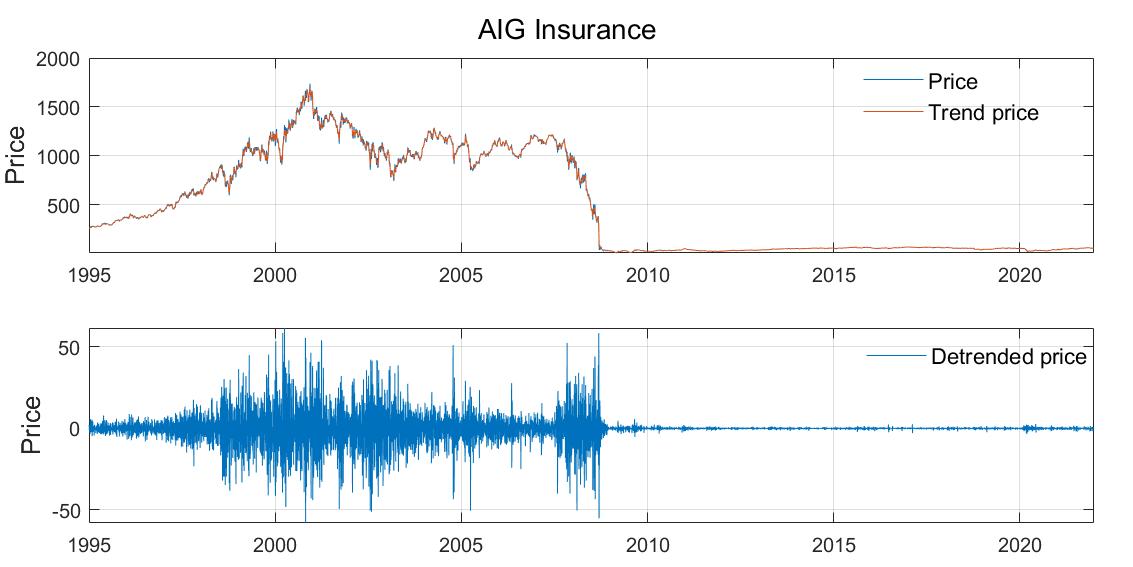}}
\end{minipage}
\caption{Exemplary price time series and their detrended dynamical signals over 1995-2021.}
\label{fig:Price}
\end{figure}

Figures \ref{fig:Price} provides an overall view, and exhibits the original price in the upper panels and the detrended price in the bottom panels of six indices/stocks. The detrended price in blue in each figure was produced by a MA filter at scale = 5, where the detrended dynamical time series were the main signals involved in the following analyses. When a crisis arises, the price becomes more dynamic, which can be clearly observed in detrended signals with the removal of the local mean. In Figure \ref{fig:Price}, the S\&P 500 and DJIA  are collections of large companies in the US market. The NASDAQ and Apple Inc. illustrate the performance of the technology industry, where Apple Inc. is the largest stock in the NASDAQ index. And the bottom panels in Figure \ref{fig:Price} represent the food industry (McDonald's) and real estate (AIG Insurance), respectively. 

Over the last 30 years, seven consecutive periods of different natures were identified based on the world economies \cite{Ref483,Ref484,Ref485}, and are marked at the top of each figure. These are:
\begin{enumerate}
    \item Economic boom/Dot-com bubble: 1997.01.01 to 1999.12.31.
    \item Internet bubble burst (Crisis-1): 2000.01.01 to 2003.12.31. 
    \item Economic recovery: 2004.01.01 to 2007.12.31.
    \item Sub-Prime mortgage crisis (Crisis-2): 2008.01.01 to 2011.12.31.
    \item Post-Global Financial Crisis (GFC) recovery: 2012.01.01 to 2014.12.31.
    \item Bull run: 2015.01.01 to 2019.12.31.
    \item COVID Pandemic (Crisis-3): 2020.01.01 to 2021.12.31.
\end{enumerate}

Observe that different crises show varying influences on different industries, which is reflected in the following complexity analysis. Apart from the impact of the crisis, the trend of price generally increases as time goes by, except the AIG insurance, which was severely attacked by the Sub-Prime mortgage crisis around 2008.

\begin{figure}[htp]
    \centering
    \includegraphics[width=\linewidth]{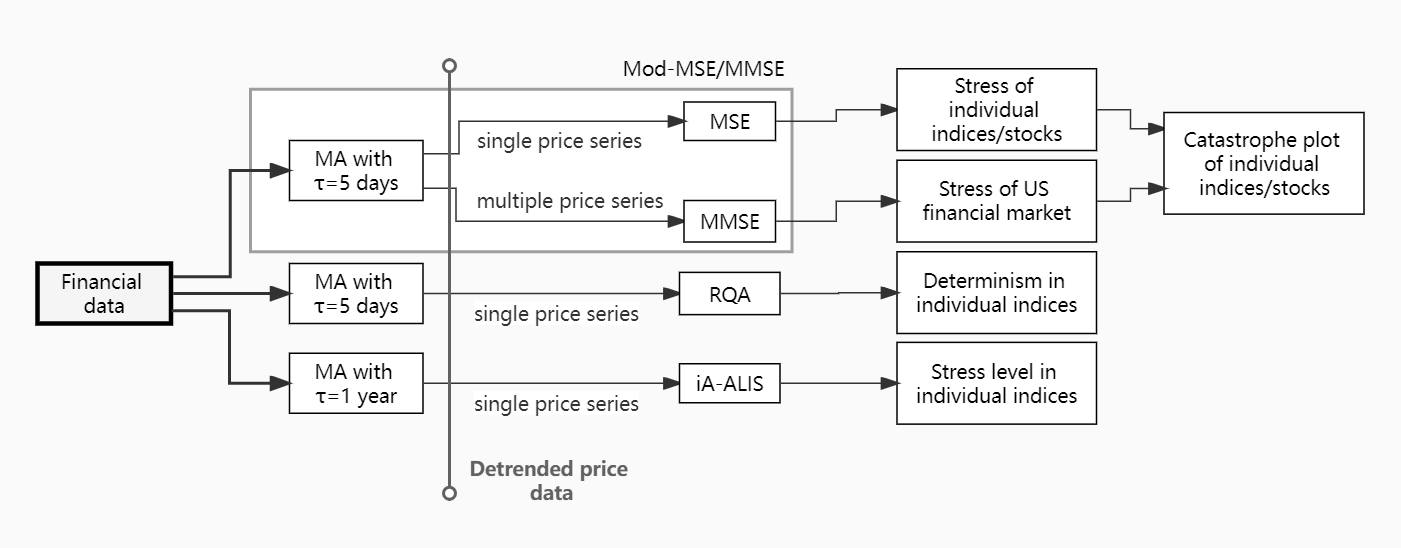}
    \caption{The employed analysis framework}
    \label{fig:framework}
\end{figure}

Due to the properties of the price time series (one data point per weekday), the length of one year contains 261 points. The analyses for univariate time series include Modified Univariate Multiscale Sample Entropy (Mod-MSE), Recurrence Plot Analysis (RQA) and Assessment of Latent Index of Stress with Instantaneous Amplitude (iA-ALIS). The analysis for multivariate time series was generated using Modified Multivariate Multiscale Sample Entropy (Mod-MMSE). Furthermore, from the Mod-MSE and Mod-MMSE analyses, the catastrophe analyzes the performance of individual price series and the overall US financial market. Figure \ref{fig:framework} shows the analysis framework using the algorithms presented in Section \ref{sec.alg}.

\section{Results and Analysis}
\label{sec.result}

\subsection{Modified Univariate Multiscale Sample Entropy (Mod-MSE) and Modified Multivariate Multiscale Sample Entropy (Mod-MMSE)}

Different settings of the pre-defined scale factors, $\tau$, have been discussed in \cite{Ref99}, where the various scale factors exhibited the same trend. Here, the short-term MA filter with $\tau$ = 5 is selected, which was able to give the most distinct tracking of the financial stress evolution. The analysis window was set as $N$ = 1044 (261 points$\times$4 years) with a 1-day increment. Therefore, given the data from 1991, the complexity plot starts from 1995, whereby each entropy value was calculated based on the history price in the past 4 years. The default parameters of Mod-MSE/MMSE were set as the embedding dimension $m$ = 2 and delay factor $l$ = 1.

\begin{figure}[htp]
    \centering
    \includegraphics[width=\linewidth]{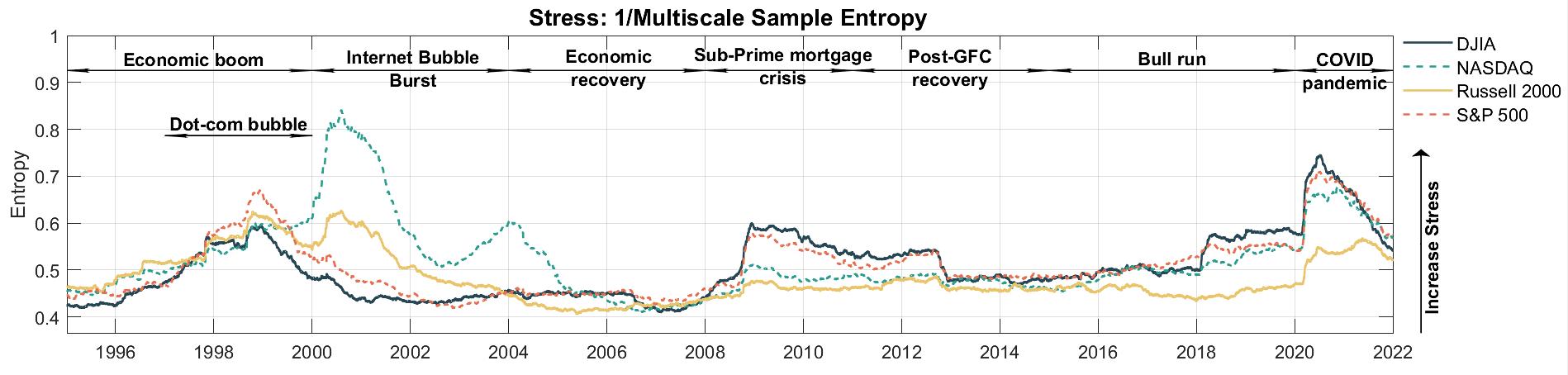}
    \caption{Financial stress of four indices (DJIA, NASDAQ, Russell 2000 and S\&P 500) estimated by Mod-MSE over 1995-2022.}
    \label{fig:mse_index}
\end{figure}

\begin{figure}[htp]
    \centering
    \includegraphics[width=\linewidth]{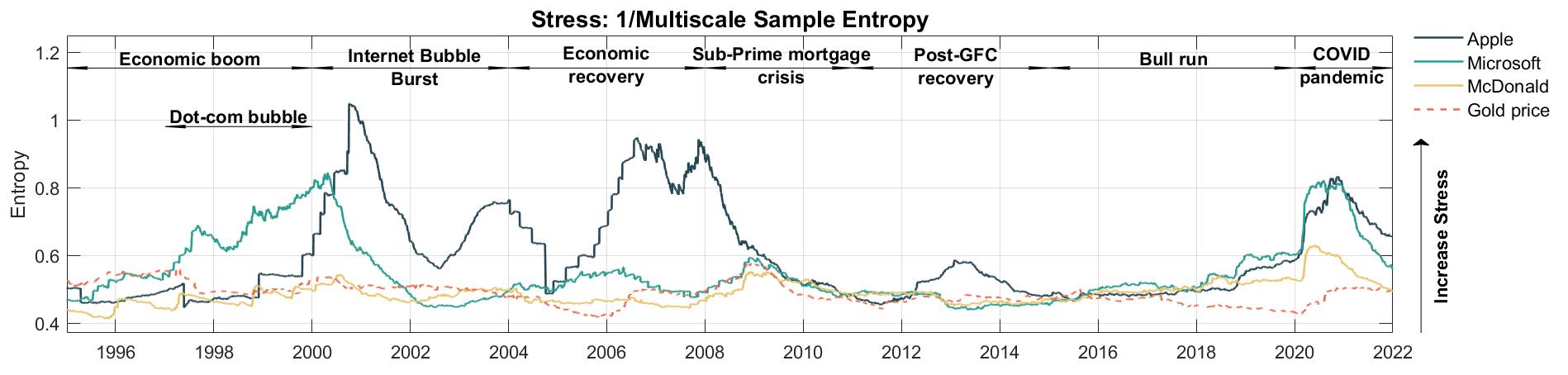}
    \caption{Financial stress of four leading equities in their sector (Apple Inc., Microsoft, McDonald's and Gold price), estimated by Mod-MSE over 1995-2022.}
    \label{fig:mse_equity}
\end{figure}

Figure \ref{fig:mse_index} and \ref{fig:mse_equity} show respectively the complexity of the 4 indices and 4 equities considered via Mod-MSE. The figures show the reciprocal of Mod-MSE, representing the level of stress according to the complexity-loss hypothesis. The increase of Mod-MSE indicates higher randomness referring to a 'normal' period and a decrease of stress when the dynamics of price are balanced and influenced by multiple factors. In terms of indices in Figure \ref{fig:mse_index}, the complexity plots validate the hypothesis that when financial crises arise (i.e., the Internet bubble burst, the Sub-Prime mortgage crisis and the COVID pandemic), all the indices showed different degrees of stress increase. As the collections of technology markets, NASDAQ (in green) showed the largest impact from the Internet bubble, while the Sub-Prime mortgage crisis had less influence. As for the COVID pandemic, all the indices illustrated a moderate increase in stress levels at the beginning of 2020 due to the general struggle of all industries. Among them, Russell 2000 took a relatively long time to recover back to normal, compared to the other three larger indices.

As for the individual equities in Figure \ref{fig:mse_equity}, observe that the most stable index was the price of gold (in red) which kept a low level of stress over time, including the COVID pandemic. The same trend can be found in McDonald's, the world's largest fast-food chain, where the stress shown by the price dynamics of McDonald's stock was as low as the gold price. Apple Inc. is the most significant individual equity in the NASDAQ index. Hence, the trend of Apple Inc. (in black) in Figure \ref{fig:mse_equity} is partly in line with the change of NASDAQ (in green) in Figure \ref{fig:mse_index}. For example, the peaks after 2000 and around 2004 are presented in both plots. Observe also the step change of Apple Inc between 1996-2006, which could be the influence of new product releases. In contrast, this effect has been reduced in the recent 10 years.

\begin{figure}[htp]
    \centering
    \includegraphics[width=\linewidth]{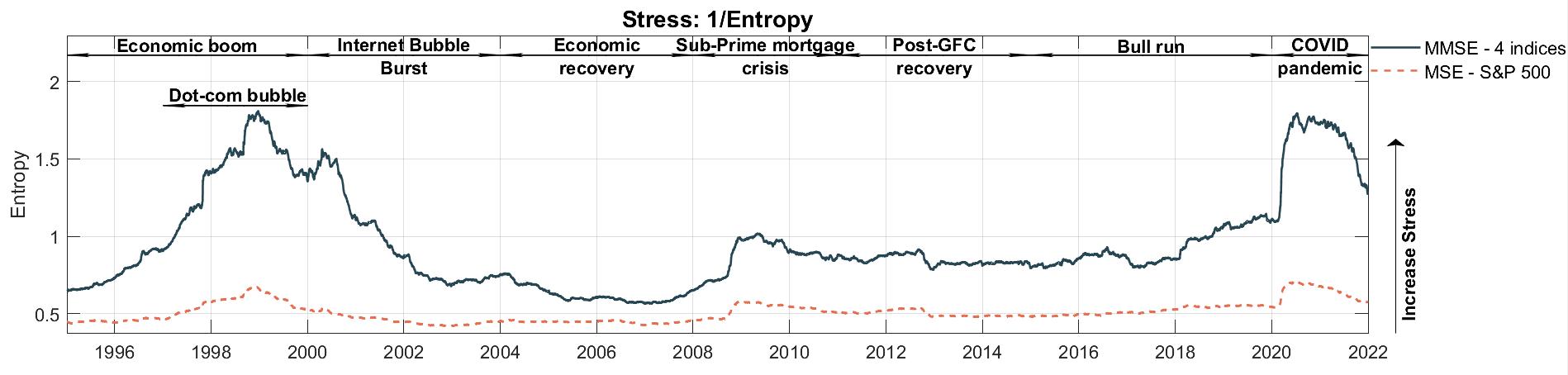}
    \caption{Financial stress of US market estimated by Mod-MSE and Mod-MMSE over 1995-2022.}
    \label{fig:MMSE}
\end{figure}

As shown in Figure \ref{fig:mse_index} and Figure \ref{fig:mse_equity}, entropy analyses implemented based on individual indices/stocks are able to assess the internal stress from the dynamics of a signal. Although S\&P 500 and DJIA are generally considered leading indices, single-channel analysis is sub-optimal to evaluate the overall performance of the US financial market. To this end, we applied the multivariate Mod-MSE with the 4 representative indices in Figure \ref{fig:mse_index} as multi-channel data. To visualize the advantages of multivariate analysis, the Mod-MSE of the leading index (S\&P 500) is jointly plotted with Mod-MMSE in Figure \ref{fig:MMSE}. Observe that during the Dot-com bubble, Sub-Prime mortgage and COVID pandemic, the stress in multi-channel increased accordingly, while for the 'normal' situations, it showed a flat stress change. With time, the stress levels of 'normal' periods increased. For example, the period of Post-GFC recovery after the Sub-Prime mortgage crisis is higher than the economic recovery after the internet bubble burst around 2006. Overall, the performance of Mod-MMSE was in line with Mod-MSE based on S\&P 500 but was more sensitive and exhibited a stronger response towards the change of environment. Therefore, Mod-MMSE was able to reflect the overall stress of the US financial market better than any individual indices/stocks.

\subsection{Recurrence Quantification Analysis}
Next, we applied the Recurrence Quantification Analysis (RQA) to yield the determinism degree (DET) of each index, which is another quantitative way to evaluate the stress level of the financial market. Recall that the EMH implies that during ‘normal’ financial regimes, stock prices behave in a random (uncertain) way \cite{Ref181}. The lower the determinism, the more stochastic components the system contains, referring to a 'normal' situation in line with the high randomness in entropy analysis. Therefore, the degree of determinism (DET) metric is consistent with the stress level of the index/stock, that is, we expect low determinism in normal situations and high determinism during the financial crisis. 

In the analysis, the pre-defined scale factor, $\tau$, was set to 5 days, and the window for RQA to 4 years. The optimal combination of the embedding dimension, $m$, and delay parameter, $l$, were jointly selected by the differential entropy-based method, proposed in \cite{Ref136}.

\begin{figure}[htp]
    \centering
    \includegraphics[width=\linewidth]{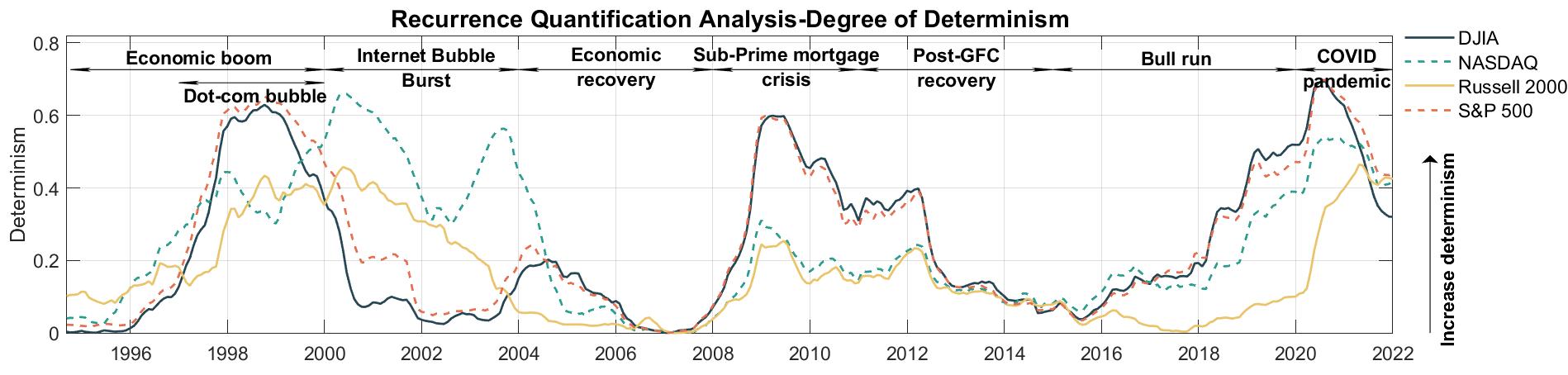}
    \caption{Degree of determinism of four indices (DJIA, NASDAQ, Russell 2000 and S\&P 500), estimated by RQA over 1995-2022.}
    \label{fig:RP}
\end{figure}

Figure \ref{fig:RP} shows the DET of 4 indices, DJIA, NASDAQ, Russell 2000 and S\&P 500. As expected, DJIA and S\&P 500 showed similar curves with high determinism during the Dot-com bubble, Sub-Prime mortgage crisis and COVID pandemic, as general responses towards environmental stress changes. On the other hand, NASDAQ in green exhibited high stress during the Internet bubble burst than the Sub-Prime mortgage crisis due to the intrinsic features of the technology market which have no relevance to housing investment. As for small companies (Russell 2000 in yellow), they showed delayed responses to crises compared to big companies, whereby the yellow line rose up following the other three indices at the beginning of the three crises, similarly, it took a long time to recover from crises. For example, around 2002, when DJIA and S\&P 500 had largely dropped, it was still at a high level; also at the end of 2021, Russell 2000 (in yellow) was still in a highly deterministic state, while the stress of other indices had mostly decreased.

\subsection{Assessment of Latent Index of Stress with Instantaneous Amplitude (iA-ALIS)}

The third methodology to estimate the stress level of financial indices/stocks is the Assessment of Latent Index of Stress with Instantaneous Amplitude (iA-ALIS) \cite{Ref141}. The ALIS was originally proposed to examine the 'economic organism' through the complexity-loss hypothesis, whereby a high-stress level is indicated by the high value of ALIS \cite{Ref99}. Here, we applied the enhanced iA-ALIS, whereby When estimating the ‘power’ of the low-frequency band and high-frequency band, the instantaneous amplitude is used in place of absolute power. Following the work in \cite{Ref99}, a four-year sliding window was employed with a 1-day increment and the detrended data was given by MA filter on a scale of one year.

\begin{figure}[htp]
    \centering
    \includegraphics[width=\linewidth]{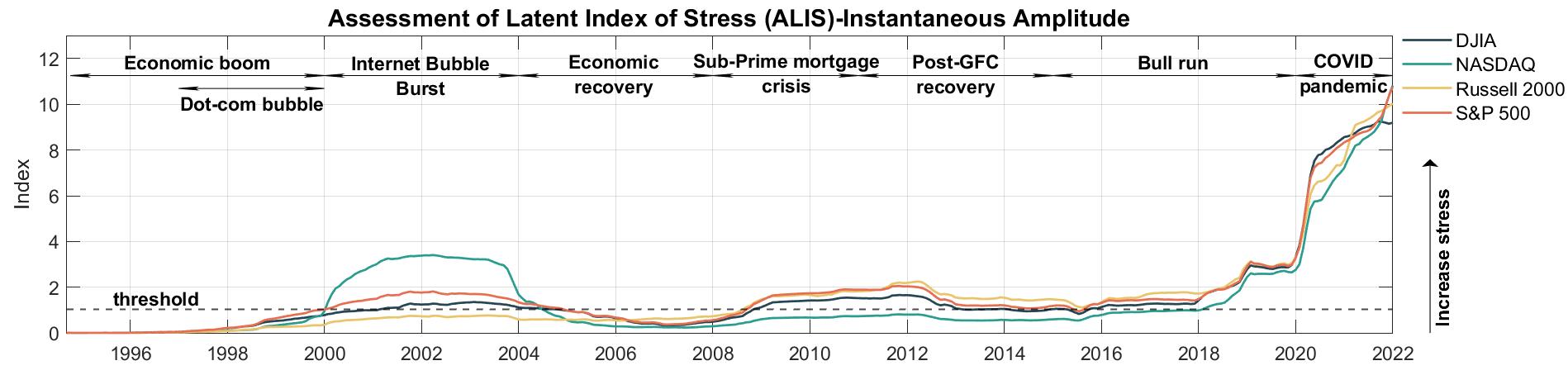}
    \caption{Financial stress of four indices (DJIA, NASDAQ, Russell 2000 and S\&P 500) estimated by iA-ALIS over 1995-2022.}
    \label{fig:ALIS}
\end{figure}

Figure \ref{fig:ALIS} shows the stress levels of every index considered (DJIA, NASDAQ, Russell 2000 and S\&P 500) given by iA-ALIS. The black dashed line in the bottom figure represents the threshold, which is the median value among all the indices over time. Observe that the iA-ALIS exhibited substantially high levels during the internet bubble burst and Sub-Prime mortgage crisis. However, when considering the impact of the COVID Pandemic, the dramatically high stress given by the global pandemic among all industries made measures of the previous two crises less significant when visualising. During the internet bubble burst, NASDAQ demonstrated a higher stress level than the other three indices as expected, which is supported by the Mod-MSE and RQA analyses.

Observe also the problem with iA-ALIS whereby the highly dynamic changes in daily price were hard to distinguish. With the same resolution, Mod-MSE in Figure \ref{fig:mse_index} and RQA in Figure \ref{fig:RP} were able to give more details of the stress evolution, while the iA-ALIS measure, as shown in Figure \ref{fig:ALIS}, evaluated the stress levels in a smooth way. Although the periods of crisis could be marked by iA-ALIS in 2000, 2008 and 2020, limited information can be observed from iA-ALIS for further analysis.

\section{Catastrophe Plots based on Entropy}
\label{sec.cata}

Complexity theory suggests that all aspects of complex problems have the characteristics of catastrophes \cite{Ref451}. Catastrophe theory is a branch of applied mathematics, developed by Rene Thom in the late 1960s \cite{Ref481}, whereby the basic idea of Catastrophe Theory aims to explain the breakdown of relationships in variables of a dynamic system \cite{Ref445}. More specifically, it offers a potential method to describe the ability of a smooth change in system parameters to generate catastrophic behaviours (i.e., abrupt, discontinuous, sudden change) in a dependent variable, termed critical points \cite{Ref445, Ref448, Ref449, Ref452}. Therefore, Catastrophe theory has the potential for describing all aspects of natural phenomena due to their complex properties; it embodies a theory of great generality which is perceived as a state of mind \cite{Ref446, Ref449}.

In the realm of physiology, Hardy \& Fazey (1987) state that physiological arousal is related to performance in an Inverted-U hypothesis when the athlete is not worried or has low cognitive anxiety. If cognitive anxiety is high, the increases in arousal pass a point of optimal arousal and a rapid decline in performance occurs \cite{Ref450}. Catastrophe Theory has also been involved in brain modelling \cite{Ref447}, however, the catastrophe model still remains in its conceptual framework state without computational analysis \cite{Ref450}. When it comes to financial applications, Catastrophe Theory represents a unique hypothesis made up of different mathematical structures, in contrast to the Efficient Market Hypothesis (EMH) \cite{Ref449}. Indeed, Catastrophe Theory has been tentatively employed to explain discontinuous jumps in bank investment \cite{Ref448} and studies have shown that it could explain the crash of the stock exchanges better than other models \cite{Ref452}. Yet, the number of studies using catastrophe theory in economics is fairly small and mostly built based on qualitative descriptions rather than quantitative applications \cite{Ref448, Ref452}. Considering the high complexity and unpredictability of the stock market and its chaotic and uncertain behaviours, Catastrophe Theory offers the potential to explain the occurrence of financial events \cite{Ref449}. To this end, we propose a practical framework for the application of Catastrophe Theory on basis of the complexity estimation through entropy.

\subsection{Arousal-Performance plot of index}

Considering the similarity between the financial and physiological systems, the analogy between catastrophe theory applied to athlete performance and index/stock performance can be drawn. The catastrophe plot in physiological systems reflects the relation between physiological arousal (anxiety) and performance. Accordingly, in the financial market, we observe the performance of individual indices/stocks evaluated by Mod-MSE; that is, the higher the stress level given by lower Mod-MSE the lower the performance. While the arousal in the physiological system is an internal trigger that determines performance, due to the intrinsic property of a financial system, we modelled the dependent factor (arousal) as the external stress imposed by the external environment on the whole market and quantified by the Mod-MMSE. Therefore, the catastrophe framework in the financial market describes the relationship between the level of overall market anxiety (external stress) and the performance of individual indices/stocks (internal stress).

\begin{figure}[ht]
    \centering
    \includegraphics[width=\linewidth]{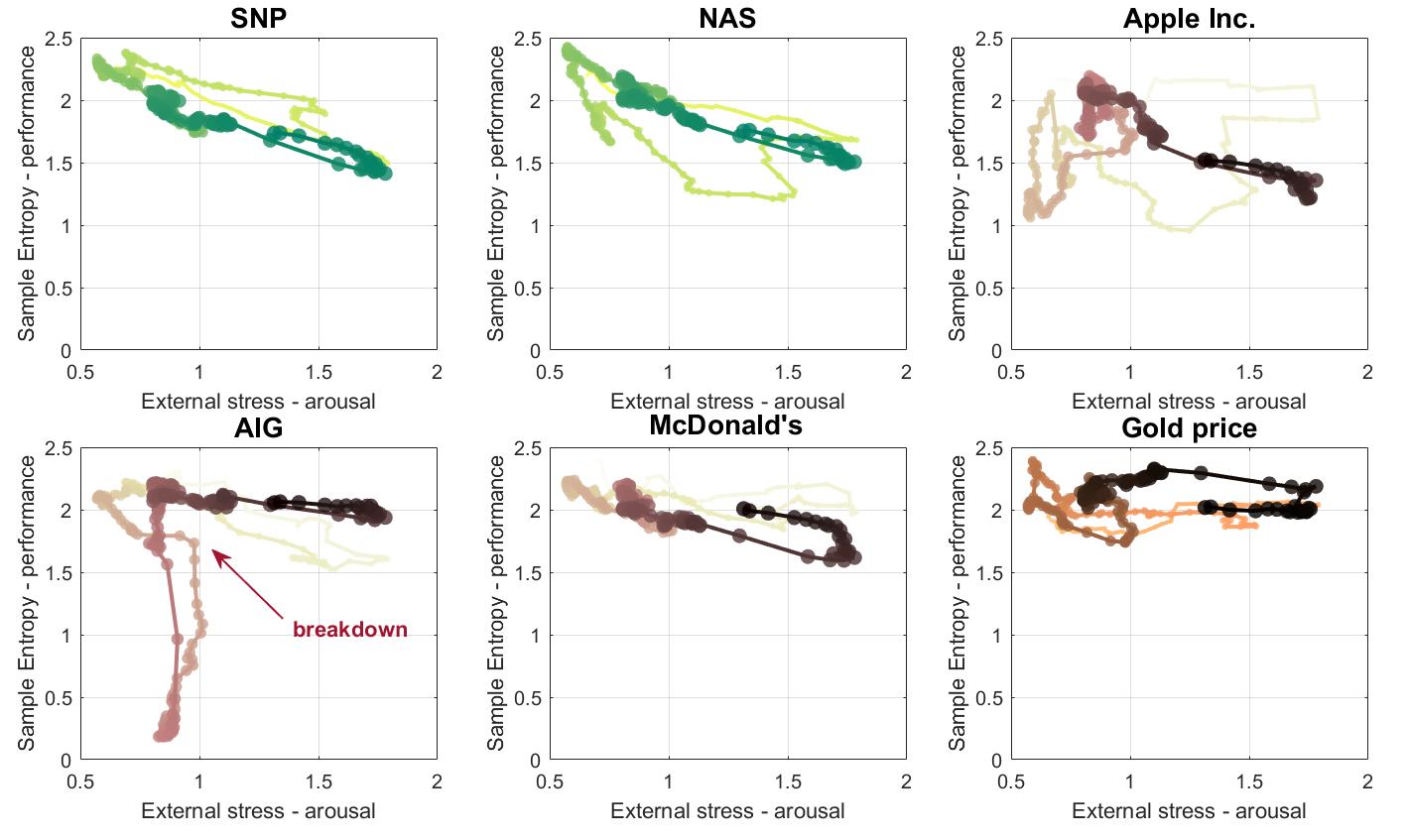}
    \caption{Evaluation of Catastrophe Plots of 6 indices (S\&P 500, NASDAQ, Apple Inc., AIG, McDonald's and gold price) over the period 1995-2022. The plots are colour-coded, starting from the lightest shade and ending in the darkest shade. The plots depict the stress of an individual asset in response to the stress of the whole market (sensitivity).}
    \label{fig:cata_all}
\end{figure}

Two indices (S\& P 500 and NASDAQ), three individual stocks in different industries (Apple Inc., AIG and McDonald's) and the price of gold (AU) were selected to demonstrate the Arousal-Performance Plot as shown in Figure \ref{fig:cata_all}. Good performance is reflected in low internal stress indicated as high Sample Entropy, while strong arousal/stimulus is designated as high external stress given by the reciprocal of Multiscale Multivariate Sample Entropy (MMSE), as discussed in Figure \ref{fig:MMSE}. The line plots in Figure \ref{fig:cata_all} depict a relationship between the performance of each index/individual stock and external stimuli during the whole 27 years (1995-2022), where light colours refer to early years and dark colours represent late years; all the plots were adjusted to the same axis scale.

The Catastrophe plots of two indices (S\&P 500 and NASDAQ) are shown in the first two panels in Figure \ref{fig:cata_all}. As a collection of big companies, S\&P 500 and NASDAQ exhibit a similar tendency of performance increase as the external stress decreases. Observe that the curve of S\&P 500 tends towards more of a linear relationship compared to the curve of NASDAQ, showing the higher predictability/regularity of S\&P 500. The larger slope of the NASDAQ curve in light green shows that the technology market was under higher stress (with sub-optimal performance) in the early years than in recent years.

Next, the curve of the most significant stock in NASDAQ, Apple Inc. is in line with the change of NASDAQ in recent years as shown in dark colours; this is due to the leading role of Apple Inc. in the technology market in the US. As for American International Group (AIG), the only individual stock that has experienced catastrophe change among the given indices is plotted in the first graph of the second row in Figure \ref{fig:cata_all}. Observe that, in the early and late years, AIG exhibits a relatively stable response to the overall external stress. What is striking is the sharp drop in the middle years (during the Sub-Prime mortgage crisis), which reflects the sharp increase of internal stress of the individual stock in response to a small change in external stress. Therefore, AIG equity exhibited two critical points in Catastrophe Plot: i) the lowest performance/highest stress level the stock could sustain; and ii) the highest performance point that could manage to bring the equity back to a normal state. Notice also that the recovering point is higher than the breakdown point. In terms of highly robust stocks such as McDonald's and gold price, both exhibit a relatively flat relationship between arousal and performance. The flat curves show the highly stable performance of the equity/index under large changes in external stress. McDonald's stock indicates a decrease in stability in recent years compared to the early years as evidenced by the more inclined tendency, while the price of gold keeps a high performance without apparent influence by the stress caused by environmental change. 

\subsection{Arousal-Performance plot of the index in crises}

\begin{figure}[htp]
    \centering
    \includegraphics[width=\linewidth]{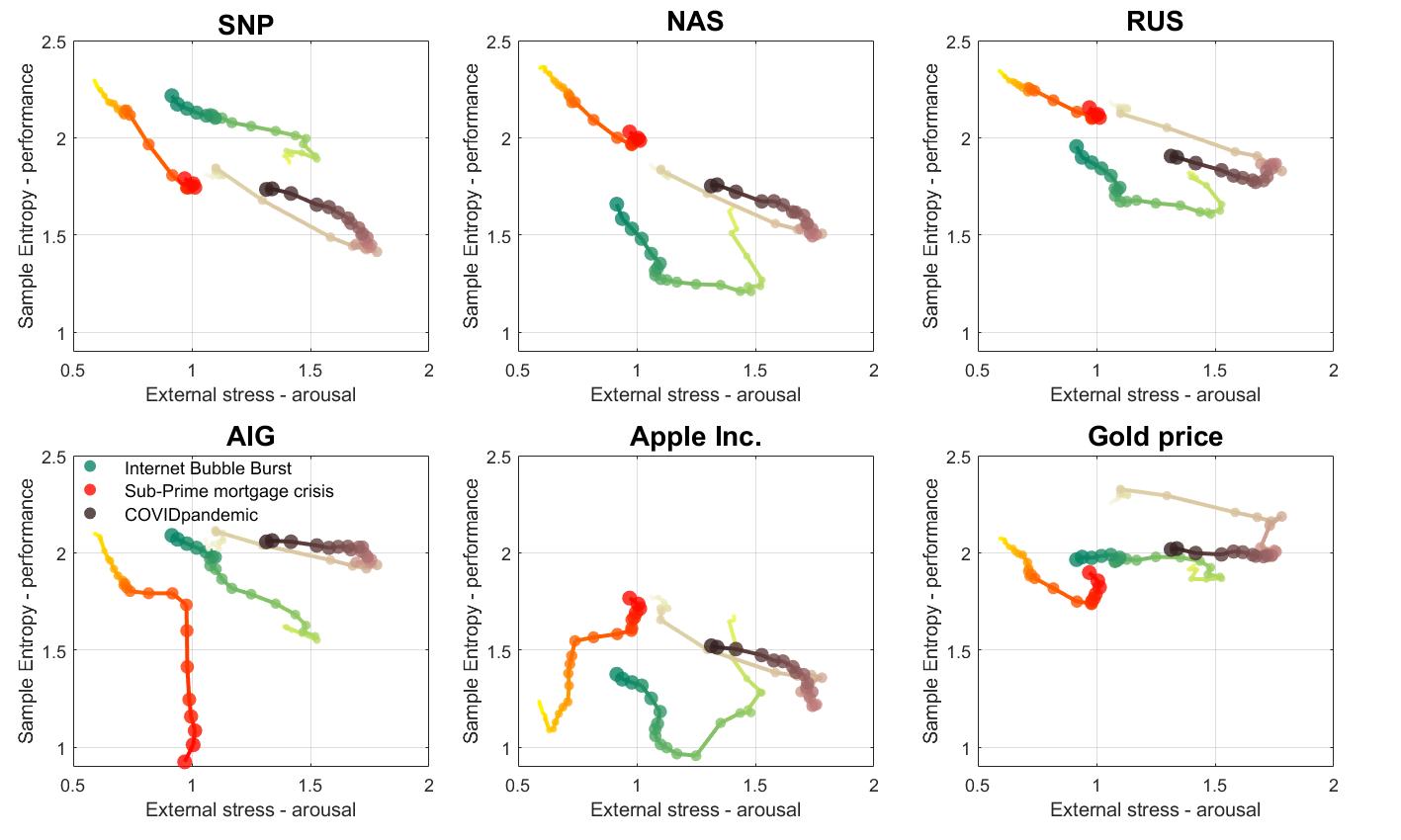}
    \caption{Catastrophe Plots of 6 indices (S\&P 500, NASDAQ, Apple Inc., AIG, McDonald's and gold price) during three selected periods of crises (Internet Bubble Burst, Sub-Prime mortgage crisis and COVID pandemic). Each segment is colour-coded, starting from the lightest shade and ending in the darkest shade lasting a period of 2 years. }
    \label{fig:cata_1}
\end{figure}

Next, we extracted the three specific 2-year periods of crisis from the previous Catastrophe Plot to discuss the response of each index/individual stock to different crises. These are the Internet Bubble Burst (IBB) between 2002-2002 in green, the Sub-Prime mortgage crisis between 2008-2010 in brown, and the COVID pandemic between 2020-2022 in red. To indicate the direction of each segment, we used the gradient colour, where the lightest colour refers to the start of the selected period of time. The selected Catastrophe Plots are illustrated in Figure \ref{fig:cata_1}.

According to the Catastrophe Theory, with a fixed increase of external stress (arousal), a good response should be shown as a small decrease or non-decrease of Sample Entropy (performance). Therefore, the slope of lines could quantitatively reflect the overall performance of the index in response to different crises. Three indices and three individual stocks are examined in Figure \ref{fig:cata_1}, with the same axis scale. The index in the first panel indicated that S\&P 500 showed stable responses to all three crises with different degrees of robustness. Observe that, among the selected segments, the Sub-Prime mortgage crisis had the largest impact on the S\&P 500 index (in red) with a sharp decrease in performance, while for NASDAQ and Russell 2000, the Internet bubble burst exhibited a larger influence than the other two crises. NASDAQ, in particular, was critically affected due to the technological domination of the index.

In terms of the equities in the second row in Figure \ref{fig:cata_1}, in line with the plot in Figure \ref{fig:cata_all}, the Catastrophe plot of the insurance company, AIG, exhibits a breaking point during the Sub-Prime mortgage crisis. It reverted back to a stable state during the COVID pandemic. As for Apple Inc. stock, the non-decreasing tendency during the Sub-Prime mortgage crisis (in red) demonstrates the limited influence of the specific stimuli (with non-decreasing performance). The most significant crisis for Apple Inc. was the Internet Bubble Burst (in green) as expected. The gold price is given in the last panel. As the most stable index, the performance of gold price has been at a relatively high level throughout three crises, with a slight drop in the Sub-Prime mortgage crisis and the COVID pandemic.

\subsection{Arousal-Performance plot of crises}

\begin{figure}[htp]
    \centering
    \includegraphics[width=\linewidth]{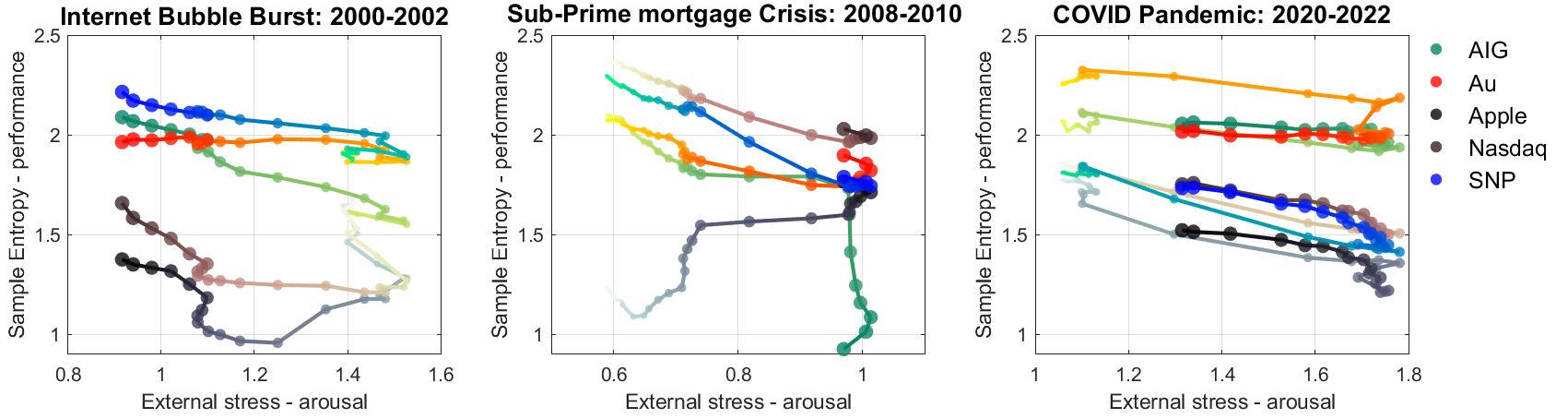}
    \caption{Catastrophe Plots of three 2-year periods of crises (Internet Bubble Burst, Sub-Prime mortgage crisis and COVID pandemic) for selected 5 indices (S\&P 500, NASDAQ, Apple Inc., AIG and gold price). Each segment is colour-coded, starting from the lightest shade and ending in the darkest shade}
    \label{fig:cata_2}
\end{figure}

Finally, to compare the performance of indices/individual stocks in each of the crises, we selected the arousal-performance plots of five indices/individual stocks in three crises separately, as shown in Figure \ref{fig:cata_2}. The left panel shows the performance during the Internet Bubble Burst between 2002-2002. Observe that the most stable two indices are the S\&P and gold prices (in blue and red, respectively) with flattening tendencies. Those which were the most significantly impacted are the NASDAQ and Apple Inc. in brown and grey. The Sub-Prime mortgage crisis in the middle graph shows that the stock that suffered most is AIG (in green) which exhibits an elbow. The remaining stocks showed similar slopes apart from Apple Inc., which exhibited an increasing performance as the external stress rises.

As for the recent COVID pandemic in the right panel, all the indices/stocks display similar responses to the crisis; that is, performance decreases at the beginning of the crisis and then gradually raises back as the stress relaxes, and the elbow point in each curve means the stock started the recovery process from the impact of the crisis. The highly aligned tendencies of all the indices emphasize the general influence of the global pandemic on all the industries in the US financial market.

\section{Conclusion}

We have estimated financial stress from the viewpoint of nonlinear dynamics and have examined the significance of complexity features in financial analysis. This has been achieved for four stock market indices and four individual equities from 1991-2021. The financial stress has been estimated based on Mod-MSE in both univariate and multivariate cases. In addition, the univariate RQA and iA-ALIS methods have been shown to give the degree of determinism and stress change, in line with the entropy-based analysis. All three nonlinear approaches have demonstrated their ability to quantify financial stress, with multivariate entropy being the most information-rich and physically meaningful.

A novel framework based on Catastrophe Theory has been proposed, where Arousal-Performance plots have been employed to visualise the response of each financial index/stock. We have adopted Mod-MMSE of four major indices as a metric of external arousal and Mod-MSE of each index/individual stock as a metric of performance. The analysis has demonstrated that the same crisis triggers different performance changes in various industries and that the same index/equity exhibits various robustness to different types of crises (the Internet Bubble Burst, Sub-Prime mortgage crisis and COVID pandemic crisis). Finally, through Catastrophe Plots, the 'performance' of index/individual stock has been qualitatively and quantitatively explored. 

The arousal and performance in this study have employed entropy-based estimation methods. Future studies will focus on incorporating different stress quantification approaches based on the proposed framework.

\appendices
\section{Univariate Multiscale Sample Entropy on Currency and Metal Price}

We have applied univariate analysis via Modified Multiscale Sample Entropy (Mod-MSE) on currency indices and metal price time series. The value of a currency can be influenced by a number of factors including the government's economic policies and their national central bank \cite{Ref98}. And national central banks are closely associated with their metallic reserves \cite{Ref482}.

\begin{figure}[htp]
    \centering
    \includegraphics[width=\linewidth]{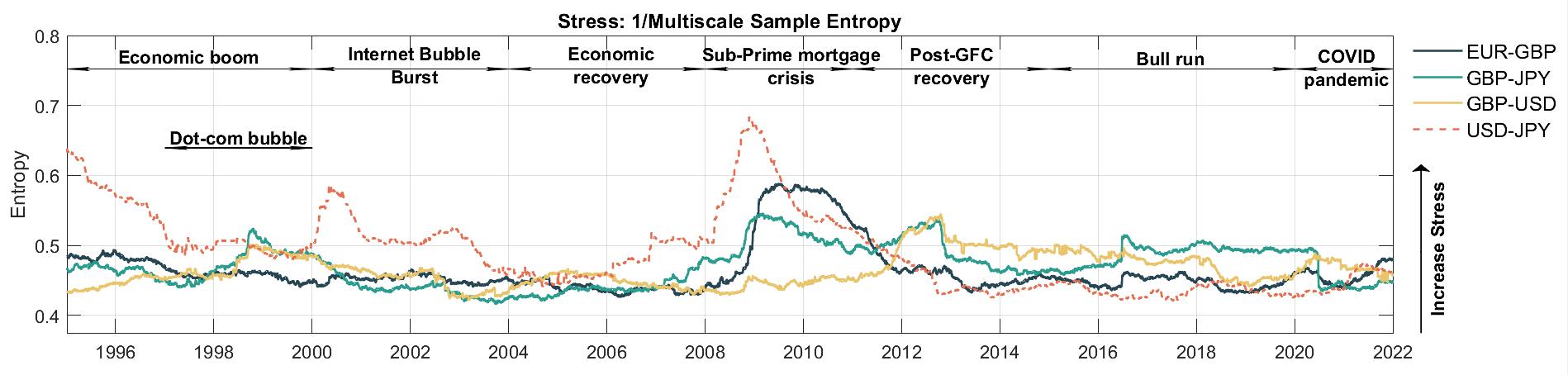}
    \caption{Financial stress of four currencies (EUR-GBP, GBP-JPY, GBP-USD and USD-JPY) estimated by Mod-MSE over 1995-2022.}
    \label{fig:mse_currency}
\end{figure}

The four currency indices are EUR-GBP, GBP-JPY, GBP-USD and USD-JPY. To demonstrate the stress level, the reciprocal of Mod-MSE is plotted in Figure \ref{fig:mse_currency} in line with the complexity-loss theory. The stress level given by Mod-MSE showed that USD-JPY (in red) was at a high-stress level before the end of the Sub-Prime Mortgage crisis. GBP-JPY (in green) and EUR-GBP (in black) exhibited higher stress levels in Sub-Prime mortgages than that in the Internet bubble burst. Generally, the Forex market was less impacted by the COVID pandemic compared to the previous two global crises.

\begin{figure}[htp]
    \centering
    \includegraphics[width=\linewidth]{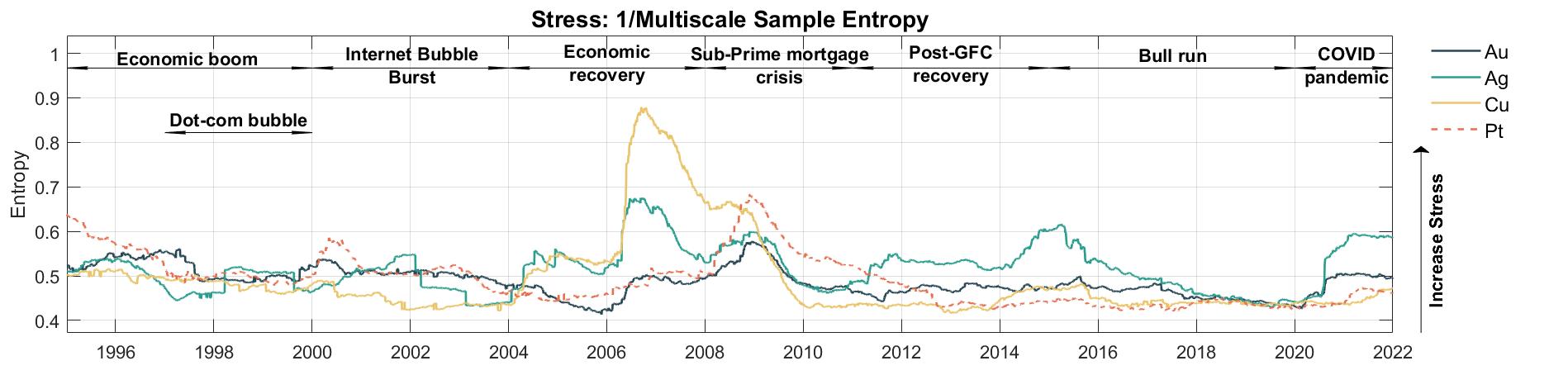}
    \caption{Financial stress of four metal prices (Gold-Au, Silver-Ag, Copper-Cu and Platinum-Pt) estimated by Mod-MSE over 1995-2022.}
    \label{fig:mse_metal}
\end{figure}

Figure \ref{fig:mse_metal} presents the stress levels of four mental prices: Gold (Au), Silver (Ag), Copper (Cu) and Platinum (Pt). In general, metal prices were less influenced by the Internet bubble burst than other stocks given in Section \ref{sec.result}. The gold price (in black) and platinum price (in red) are at a low-stress level over time as expected. The copper price has shown the most sensitivity during Sub-Prime mortgage crisis, while silver received more impact from the attack of the COVID pandemic. 

\section*{Acknowledgment}

The authors would like to thank...

\ifCLASSOPTIONcaptionsoff
  \newpage
\fi

\bibliographystyle{ieeetr}
\bibliography{bibtex/Ref}

\end{document}